# Multiplexed Readout for 1000-pixel Arrays of Microwave Kinetic Inductance Detectors

Joris van Rantwijk, Martin Grim, Dennis van Loon, Stephen Yates,
Andrey Baryshev and Jochem Baselmans

*Abstract*—Microwave Kinetic Inductance Detectors (MKIDs) are the most attractive radiation detectors for far-infrared and sub-mm astronomy: They combine ultimate sensitivity with the possibility to create very large detector arrays, in excess of 10 000 pixels. This is possible by reading-out the arrays using RF frequency division multiplexing, which allows multiplexing ratios in excess of 1000 pixels per readout line. We describe a novel readout system for large arrays of MKIDs, operating in a 2 GHz band in the 4–8 GHz range. The readout, which is a combination of a digital front- and back-end and an analog up- and down-converter system, can read out up to 4000 detectors simultaneously with 1 kHz data rate. The system achieves a readout noise power spectral density of −98 dBc/Hz while reading 1000 carriers simultaneously, which scales linear with the number of carriers. We demonstrate that 4000 state-of-the-art Aluminium-NbTiN MKIDs can be read out without deteriorating their intrinsic performance.

## I. INTRODUCTION

THE next generation instruments for mm wave and sub-mm astronomy will require very large arrays of background limited radiation detectors operating at wavelengths between 3 mm and 0.03 mm (frequencies between 0.1 and 10 THz). New ground based observatories such as CCAT [1] need imaging instruments with close to a million pixels to fill the telescope field of view. New spectroscopic instruments based upon direct detectors, such as DESHIMA [2] and SuperSpec [3] require about 1000 pixels to obtain a single spectrum for a single source. Even space based instruments, such as SPICA-SAFARI, are planning instruments with of the order of 10 000 pixels. Microwave Kinetic Inductance Detectors (MKIDs), pioneered in 2003 [4], have become the detectors of choice for these instruments due to their intrinsic capability of frequency division multiplexing at microwave frequencies. Both antenna coupled MKIDs [5] and lumped element MKIDs [6][7] have shown background limited sensitivities combined with a high radiation coupling efficiency. However, a compact, highly integrated readout system with a large bandwidth capable of reading out in excess of 1000 MKIDs without deteriorating their intrinsic performance is not available. This is especially problematic for aluminium based antenna coupled MKIDs which have shown the best sensitivities [8] but operate at relatively high readout frequencies of 2–8 GHz. In this paper we describe an analog/digital readout system with a 2 GHz bandwidth, operating in a 4–8 GHz frequency range, capable of reading out up to 4000 MKIDs at a kHz data rate. Similar systems for multiplexed readout of MKID arrays are already in operation [9][10][11]. Compared to those systems, our readout offers a unique combination of speed, bandwidth, number of multiplexed channels and the capability to read-out ultra-sensitive detectors optimized for spectrometric or even space applications.

## II. MKID ARRAYS

The concept of frequency division multiplexed readout using MKIDs is explained in Fig. 1. A MKID can be defined as a superconducting resonance circuit, resonating at a few GHZ, capable of absorbing sub-mm radiation. A large array of MKIDs can be constructed by coupling many MKID resonators to a single transmission line. Each MKID is designed to have a slightly different resonance frequency at which it will be read-out. The forward transmission of the common transmission line, S21, is shown for a large MKID array in Fig. 1a. Radiation absorption in an MKID results in a reduction in its resonance frequency and a decrease in its Q factor (i.e. an increase in the resonator width). This is indicated in Fig. 1b. A single frequency readout tone at the original resonance frequency of each resonator (indicated by the black lines) allows the measurement of the change in the complex transmission S21 and thereby the measurement of the amount of absorbed radiation power. Note that the THz

Manuscript submitted June 17, 2015. This work was supported as part of a collaborative project, SPACEKIDS, funded via grant 313320 provided by the European Commission under theme SPA.2012.2.2-01 of Framework Programme 7.

Joris van Rantwijk, Martin Grim, Dennis van Loon and Stephen Yates are with the SRON Netherlands Institute for Space Research, Utrecht/Groningen (email: J.F.van.Rantwijk@sron.nl; W.M.Grim@sron.nl; D.van.Loon@sron.nl; S.Yates@sron.nl).
Andrey Baryshev is with the SRON Netherlands Institute for Space Research, Groningen and with the Kapteyn Astronomical Institute, University of Groningen, The Netherlands (email: A.M.Baryshev@sron.nl).
Jochem Baselmans is with the SRON Netherlands Institute for Space Research, Utrecht, and with Delft University of Technology, The Netherlands (email: J.Baselmans@sron.nl).

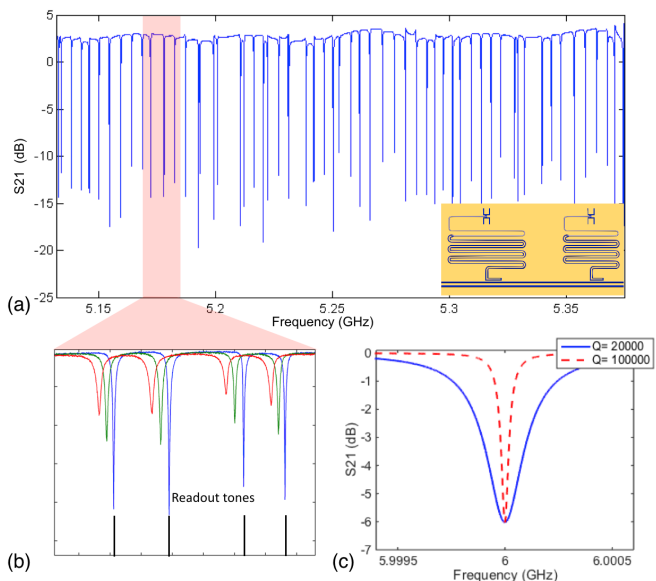

Fig. 1. MKID array readout concept. (a) Transmission of a large array of MKIDs. Each resonance feature corresponds to a different MKID. The inset shows 2 MKID detectors from the mask design of the lens array in [17]. The detectors have different resonance frequencies, set by their length. (b) Detail of panel a where we show the effect of increasing sub-mm power falling on the detectors: The resonances get broader and shallower and move to lower frequencies. The black lines indicate the placement of readout tones. (c) Low Q MKID, similar as in panel a and b, as used for ground based imaging system. The much narrower high Q resonance is typical for spectrometer applications on the ground or imaging from space, and allows for much more pixels per unit of bandwidth.

frequency band for which the device is sensitive is independent of the microwave readout frequency.

The Q factor of an MKID, is determined by the losses caused by the radiation power absorbed in the detector. For a sub-mm camera operating from the ground, these losses are determined under nominal operation by the background loading power due to the warm atmosphere, which is of the order of 10–100 pW. This results in Q ~ 10 000–50 000.

For spectrometer applications or space based imaging instruments, the power per pixel is much lower: 0.01–1 pW, resulting in much higher Q factors. This allows higher multiplexing ratios as shown in Fig. 1c.

### III. SYSTEM REQUIREMENTS AND ARCHITECTURE

The system described in this paper is envisioned for Aluminium based MKIDs. As such, it must fulfill the following basic requirements:
1) Operating in a 4–8 GHz band;
2) Readout noise below −91 dBc/Hz, i.e. significantly less than the intrinsic MKID output phase noise [5][8];
3) A large bandwidth, which we have chosen to be 2 GHz due to availability of commercial electronic parts with the required noise performance;
4) Ability to read-out more than 1000 MKIDs without degradation of the intrinsic MKID sensitivity;
5) Ability to read-out high-Q resonators (Q ~ 100 000) suitable for future ground- and space based systems.

The architecture of our readout system is illustrated in Fig. 2. A pair of digital-to-analog converters (DACs) generate a complex in-phase/quadrature-phase (I/Q) signal in an intermediate frequency (IF) band from −1 to +1 GHz. The IF signal is converted to the readout frequency (RF) by I/Q-mixing with a reference signal from a local oscillator (LO); effectively shifting the tones to a 2 GHz band centered around the LO frequency. This band can be placed anywhere in the 4–8 GHz range by an appropriate choice of LO frequency between 5–7 GHz. The RF tones, each tuned to the resonance frequency of a single detector, pass through the common feedline of the MKID array. Each MKID modulates the amplitude and phase of its designated carrier tone depending on the amount of absorbed radiation. The modulated carriers are then converted back to IF by a second mixer, operating at the same LO reference. The resulting complex I/Q signal is sampled by a pair of analog-to-digital converters (ADCs), and the amplitude/phase modulation of the carrier tones are extracted through digital signal processing.

Programmable attenuators in the RF path make it possible to adjust the signal power at the MKID array while still using the full dynamic range of the DACs and ADCs

The homodyne nature of our system, mixing up and down with a common LO, makes it less sensitive to correlated noise caused by phase- or amplitude noise in the LO signal. This is especially important because we are recording a narrow bandwidth around each carrier; any mismatch between the two mixers would complicate the task of demodulating the attenuated carriers.

The high-level system requirements translate to design parameters: 2 GHz bandwidth corresponds to a sample rate of 2 GS/s for each DAC/ADC. The noise requirement drives the dynamic range specification of the DAC and ADC. The need to tune carrier frequencies to high-Q resonators implies high frequency resolution, which drives the length of waveform memories in the digital part of the system. The following sections present a detailed design.

### IV. DIGITAL ELECTRONICS

This section describes the digital part of the readout system. Its main components are a DAC board for carrier generation, an ADC board for data acquisition, and a fast Fourier transform engine (FFT) for demodulation.

#### A. Carrier Generation

Carriers are generated in the form of a complex IQ signal in the IF band from −1 to +1 GHz. The carrier generator is composed of a Virtex-7 FPGA board and an FMC daughter board equipped with two AD9129 14-bit DACs clocked at 2 GS/s. The two DACs represent the in-phase and quadrature components of the IF signal.

The I and Q waveforms are created by placing a set of carriers on a grid of $2^{19}$ frequency bins and subsequently computing an inverse fast Fourier transform. This method supports an individual choice of amplitude for each carrier, which is desirable because the MKIDs in an array may require slightly different readout power levels. The resulting waveforms are cyclic with a period of $2^{19}$ samples. These are stored in internal RAM in the FPGA. The length of the waveforms determines the frequency resolution: all carrier frequencies in our system are multiples of 3.8 kHz (2 GHz sample rate divided by $2^{19}$ samples). As an advantage of the limited frequency resolution, any distortion products at the DAC output can only exist at discrete frequencies, not inside the signal bandwidth around the carriers. Note that the bandwidth of a Q=100 000 resonator at 4 GHz is 40 kHz, i.e. we can place a readout tone with a precision given by the resonator bandwidth divided by 10.

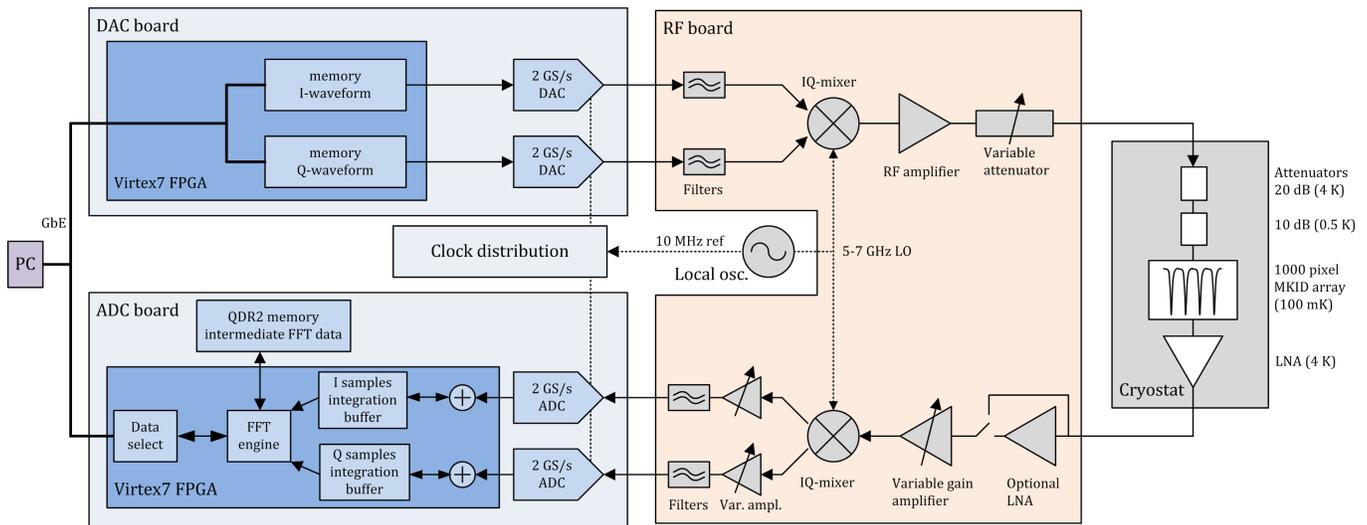

Fig. 2. System architecture including digital electronics (blue), RF electronics (red) and cryostat with MKID array (grey).

An analog reconstruction filter is typically needed to remove high-frequency components that could alias during data acquisition. The AD9129 DAC is capable of running at twice the nominal conversion rate while applying an internal digital interpolation filter. The filter provides 40dB attenuation above the Nyquist frequency and thus relaxes the requirements on the reconstruction filter without sacrificing useful bandwidth.

To maximize the signal-to-noise ratio (SNR) of the waveform, it is necessary to use the largest possible carrier amplitude without exceeding the range of the DAC. This implies that the crest factor (the ratio between peak amplitude and RMS amplitude of the signal) must be as low as possible. We currently assign random initial phases, resulting in crest factors between 12 dB and 14 dB for waveforms with up to 4000 carriers. For this system this is the most practical solution: Since the system SNR is limited by the ADC (see Table I), and the MKIDs add a quasi-random phase to the readout tones, it is virtually impossible to get better crest factors, event with advanced techniques such as presented in [13][14].

*B. Data Acquisition and Demodulation*

The IF carrier signal is upconverted to RF, passed through the MKIDs, then downconverted back to a complex I/Q signal in IF (see section V below). The resulting IF signal contains the same carriers as generated by the DACs, except the amplitude and phase of each carrier are now modulated by the response of a MKID. The I and Q components are sampled by a dual-channel, 10-bit ADC clocked at 2 GS/s. We use an EV10AQ190 ADC, located on an FMC daughter board and attached to a Virtex-7 FPGA board.

The FPGA firmware takes the I and Q sample streams from the ADC and accumulates frames of $2^{19}$ samples in separate overlap-add buffers. Each new frame is added, sample-by-sample, to the intermediate values in the buffer until 24 frames have been processed. The accumulated frame is then sent to the FFT engine and the overlap-add process restarts from zero. The overlap-add buffer is located in internal RAM in the FPGA. Double buffering must be used to allow accumulation of a new frame while the previous frame is being transferred to the FFT engine, at a rate corresponding to 24 frames of $2^{19}$ points at 2 Gsample/s, i.e. 159 frames/s. We calculate a $2^{19}$-point complex Fast Fourier Transform (FFT) for each accumulated frame. A mixed-radix algorithm is used to compute the $2^{19}$-point FFT as a series of $2^9$-point FFTs, followed by phase rotations, a $2^9 \times 2^{10}$ matrix transposition and finally a series of $2^{10}$-point FFTs [15]. The small FFTs are implemented as standard blocks from the Xilinx library. The phase rotation is implemented as a Xilinx CORDIC block. Transposing a frame of $2^{19}$ elements requires more RAM than is available inside the FPGA, hence an on-board QDR2 SRAM memory bank is used to store the frame during the transpose step. The FFT logic is clocked at 125 MHz. Internal calculations use 24-bit fixed point numbers, with a configurable scaling schedule in the FFT blocks to avoid numeric overflow.

The output of the FFT engine consists of frames of $2^{19}$ complex bins at a rate of 159 frame/s. Only a subset of the bins corresponds to carrier tones. We implement a bin selector to extract only those bins (configurable up to 4096 bins) and discards the rest of the bins. The selected bins are truncated to 16-bit signed numbers and sent to the PC via Ethernet. The bin selector reduces the output data rate to 21 Mbit/s.

Our data acquisition system effectively provides a time series of the complex amplitude of each carrier at a readout rate of 159 Hz. An alternative configuration is supported where the frame size is $2^{16}$ points with a readout rate of

TABLE I
DAC AND ADC SPECIFICATIONS

|  | DAC | ADC |
|---|---|---|
| Model | AD9129 | EV10AQ190 |
| Manufacturer | Analog Devices | E2V |
| Resolution | 14 bits | 10 bits |
| Sample rate | 2 GS/s | 2 GS/s |
| Effective number of bits | - | 7.9 bits |
| SNR (max sine wave) | - | 49 dB |
| Noise spectral density (n=1) * | −165 dBc/Hz | −142 dBc/Hz |
| Noise spectral density (n=1000) ** | −121 dBc/Hz | −98 dBc/Hz |

*) Noise spectral density relative to a full scale complex carrier, calculated as −49 dBc − 10×log$_{10}$(2 GHz) = −142 dBc/Hz

**) Noise spectral density relative to a carrier 44 dB below full scale, typical for 1000 tones assuming 14 dB crest factor, calculated as −142 dBc/Hz + 10×log$_{10}$(1000) + 14 dB = −98 dBc/Hz

1272 Hz. That configuration makes it possible to measure the MKID signals with a larger bandwidth, at the cost of limiting the choice of carrier frequencies to multiples of 30.5 kHz. The increased readout rate enables observing interactions between cosmic rays and the MKID array, given the intrinsic lifetime of ~ 1 ms for aluminium MKIDs. The two configurations have equal readout noise spectral density. Measurements presented in this paper were taken at the 159 Hz frame rate.

Table I shows the specified performance of the DAC and ADC. Sample rate and noise spectral density are the primary performance criteria for our application; digital word length is of minor importance. From this perspective, both selected parts are among the best commercially available converters. It can be seen that the ADC is a dominant noise source, producing a noise spectral density of −98 dBc/Hz for 1000 carriers.

*C. Digital Readout Integration*

The DAC and ADC boards both require a 2 GHz sample clock. We use a custom clock distribution board to generate a single 2 GHz signal and distribute it to both converters. The use of a common sample clock ensures that the DAC and ADC convert at the exact same sample rate, which is necessary to align each carrier with a single FFT bin. In addition, close-in phase noise in the clock source will be highly correlated between the DAC and ADC clocks. The correlated part of the noise will thus be canceled in the final demodulated signal.

The clock board is based on an AD9520 clock generator with a separate on-board voltage-controlled oscillator, which achieves lower phase noise than the internal oscillator in the AD9520. The clock runs in phase-locked loop with an external 10 MHz reference derived from the LO. White phase noise in the sample clock is the major contributor to the total RMS jitter. In a system with single carrier frequency $\omega_c$ and jitter $\sigma_t$ the SNR determined by the clock alone equals $20\times\log_{10}(\omega_c \sigma_t)$. In our multicarrier system we can make a worst case estimate for the total allowed jitter by assuming a carrier frequency of 1 GHz and a minimum SNR of 54 dB for the clock alone, which results in a maximum jitter of 300 fs$_{rms}$. Combined with the clock source of 2 GHz, the white phase noise in the clock signal should be below 148 dBc/Hz. Close-in phase noise will be attenuated because it correlates between DAC and ADC. However, high-Q filters or MKIDs in the signal path can reduce this correlation. Simulations of the MKID transfer function indicate that close-in phase noise must be less than −50 dBc/Hz at 1 kHz in order to stay below the intrinsic phase noise of the MKID. This is easy to obtain from a 10 MHz reference source; our system has −80 dBc/Hz phase noise at 1 kHz offset from the carrier, i.e. 30 dB margin.

The two FPGA boards with DAC and ADC daughter boards, as well as the clock distribution board are mounted in a 3U 19-inch subrack as shown in Fig. 3. The rack also contains power supplies and forced air cooling. The sample clock is distributed from clock board to the converters via short semi-rigid coax cables. The FPGA firmware is controlled by software running on a Linux PC. The boards communicate with the PC via Ethernet, using an application-specific protocol based on raw packets. An Ethernet switch board is mounted in the rack to provide a 1 Gbit fiber link to the controlling PC and 1 Gbit backplane links to the two FPGA boards. The PC is responsible for further analysis and visualization of data, as well as monitoring of temperatures and voltages in the electronics.

V. RF ELECTRONICS

The RF electronics consist of an upconverter chain, connected between the DAC ouput and MKID cryostat input, and a downconverter chain connected between the MKID cryostat output and the ADC. The system enables tuning of the carrier power at the MKID level by 30 dB, to support readout of the MKIDs at their optimal level. The entire RF system is implemented on a single 160x100 mm printed circuit board with RF drop-in components on the board surface and all control and bias wiring in buried layers. The board is a 6-layer design with dielectrics composed of FR-4 and RO4350B. Large copper thermalisation structures are used to efficiently cool the card. The final board is shown in Fig. 4, a block diagram is shown in Fig. 2.

*A. Upconverter*

The upconverter is based upon an IQ mixer (Hitite HMC525LC4), where the I and Q port are each connected to a single DAC. We use low pass filters (Mini-Circuits LFCN-1000) to reject spurious from the DAC. We make the connection to the mixer I and Q port using a 100 nF capacitor in parallel with a 49.9 Ω resistor to ground. This ensures ac coupling and a good IF match since the IF ports of the mixer are not matched to 50 Ω. The advantage of using an IQ modulator is that it enables the correction of the phase and amplitude imbalances of RF hybrid inside the mixer by changing the phase of the I and Q input signal (see section VI) The LO signal is generated by a commercial, low noise synthesizer (Agilent E8257D) and is split between the up- and downconverter channel using a 3dB power divider (GP2X1). We use a gain block (HMC619LP5) to reach +17 dBm LO power required for maximum gain flatness over the entire frequency band. After the mixer we us a gain block consisting of two amplifiers (HMC619LP5, +11 dB) separated by a 3 dB attenuator to amplify the RF signal to a peak envelope power (PEP) ≤ 7.7 dBm. A step attenuator (HMC425LP3) is used as variable attenuator to tune the readout signal to the required power level for the MKIDs. The maximum PEP at the upconverter output is +15 dBm and can be attenuated in 0.5 dB steps down to −15 dBm. For 1000 tones this corresponds to −40 to −70 dBm single tone power (assuming crest factor C=14 dB).

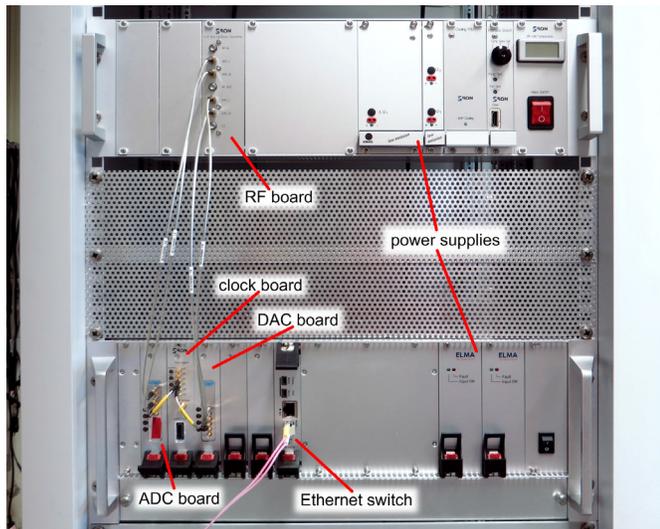

Fig. 3. Integrated digital readout (bottom) and RF electronics (top) mounted in a 19" rack. IF signals are connected via semi-rigid coax cables.

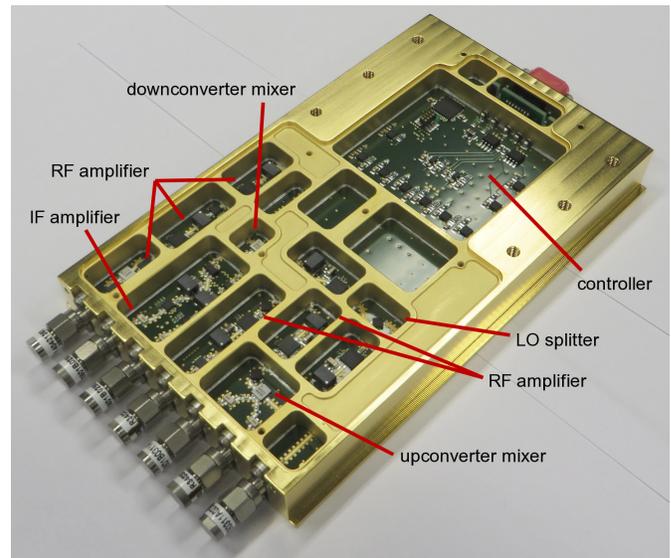

Fig. 4. RF upconverter/downconverter electronics board.

*B. Cryostat*

MKIDs are typically operated at 100 mK using a cryogenic system equipped with a low noise amplifier (LNA) at a temperature of 4 K to create the lowest possible system noise. Such an amplifier has an input noise temperature of ~4 K [18] and a gain of ~40 dB. The phase and amplitude noise power spectral density (PSD) of the cryogenic system are thus both identical and determined by the LNA noise temperature. The readout signal power for each MKID is typically between −75 and −105 dBm, resulting in a PSD in between −116 and −87 dBc/Hz. To reach these noise levels, the output noise of the upconverter chain must be low enough, which can only be achieved if we use ~30 dB of cold attenuation before the MKID chip to eliminate 300 K thermal noise. It is important to realize that a typical cryogenic LNA has an input 1 dB compression point $P_{1dB}$ ~ −40 dBm. A 1000 tone readout signal with a single tone power of −90 dBm results in PEP= $P_{tone} \times n_{tones} \times C_{crest}$=−46 dBm (with $C_{crest}$=14 dB). This illustrates that the LNA and array design must be an integral part of the system design. Note that the transmission of the cryostat is between 0 and −10 dB, including LNA, 30 dB attenuation and cable losses. This is taken as a design parameter for the downconverter.

*C. Downconverter*

We use two gain blocks (HMC619LP5) with a step attenuator (HMC425LP3) in between to create a variable gain amplifier as indicated in Fig. 2. We use an additional low noise gain block (HMC772LC4) that can be switched in or out of the signal path using two switches (HMC232LP4). The extra LNA is needed to create 10 dB extra dynamic range compared to the upconverter to compensate for frequency dependent losses in the RF cabling. The result is that we drive the IF input ports of the IQ mixer always with the same power which makes system calibration easier. The mixer (HMC525LC4) is the same as the upconverter and also connected to the LO and IF signals in the same way. The only addition is that we use amplifiers (HMC470LP3) in the IF lines to the I and Q ports of the ADC to be able to reach the full input power range of the ADC.

VI. PERFORMANCE MEASUREMENTS

We discuss three performance aspects of our readout system: sideband rejection, intrinsic noise of the readout electronics with RF loopback, and noise of the full system with cryostat and MKID array. All measurements are done with the integrated readout system, consisting of digital electronics as described in section IV and RF electronics as described in section V.

*A. Sideband Rejection*

To represent a single complex carrier as an IQ signal, the I and Q components must have exactly equal amplitude and 90 degrees phase separation. Any mismatch in amplitude or phase causes sideband leakage. This is problematic because the leaked power passes through the cryostat, unaffected by any MKID resonance, then gets partially demodulated with the carrier. We suppress sideband leakage by adjusting the amplitude and phase of the DAC waveforms to compensate for mismatches in the IF path and RF hybrid in the IQ mixer. We use a specific calibration procedure to find near-optimal gain/phase adjustments in a small number of steps, faster than the method presented in [16]. Calibration starts with an initial estimate of the amplitude mismatch, obtained by measuring the gain from the digital I waveform to the RF output and separately the gain from the digital Q waveform to the RF output. Compensation for the initial amplitude mismatches are then applied to the 1000 carriers and an adjusted IQ waveform is sent to the DAC. We subsequently measure the sideband leakage and assume that this leakage is due to phase mismatch, since amplitude mismatch has already been compensated. We calculate the amount of phase mismatch that corresponds to the measured leakage. Since the sign of the phase mismatch is unknown, a phase compensation is attempted in both directions followed by measurements of the resulting leakage, and the best phase compensation is retained. This cycle is repeated two more times to find successively improved phase compensation. Two iterations of a similar adjustment routine are then performed to improve the amplitude compensation.

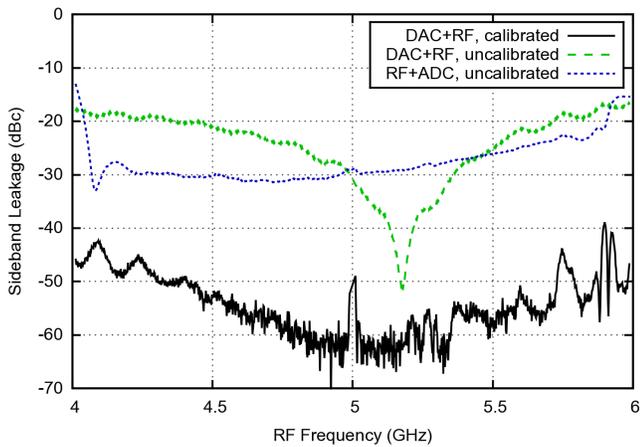

Fig. 5. Sideband leakage of digital+RF electronics: DAC+upconverter before and after calibration (measured with spectrum analyzer on RF output), and ADC+downconverter without calibration (measured via RF loopback).

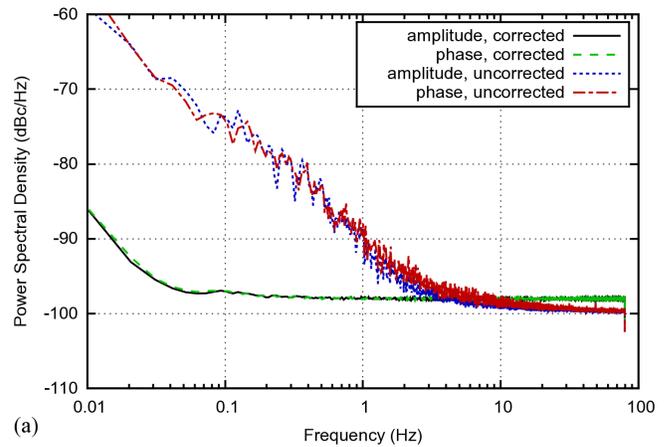

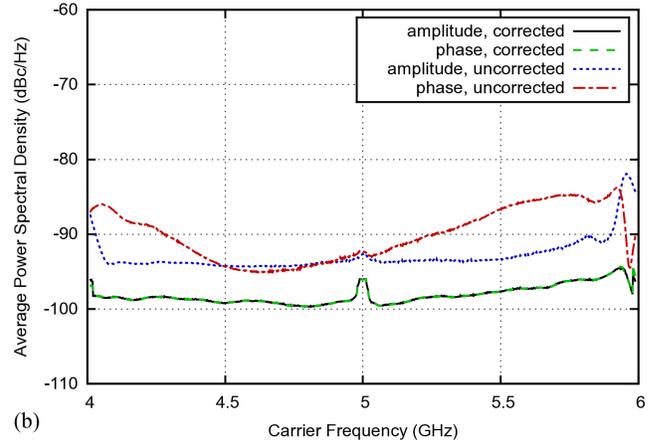

Fig. 6. Noise spectral density of demodulated carriers, measured with 1000 tones in RF loopback configuration without MKIDs, 159 frame/s, 5 GHz LO frequency. (**a**) Noise spectral density of demodulated signal, averaged over all carriers, either amplitude or phase readout, with or without blind-tone correction, measured for 100 seconds with 7 averages. (**b**) Noise level as function of carrier frequency, averaged over 0.01 - 80 Hz band.

The result of our IQ calibration is displayed in Fig. 5. The calibrated system has a typical sideband leakage of −50 dB, versus −20 dB for an uncalibrated system.

The downconverter and ADC are not calibrated for sideband leakage. While possible in principle, this would require the FFT firmware to record two bins for each carrier: the carrier bin itself and its sideband image. However, the calibrated upconverter and nominal downconverter together give a spurious signal on each demodulated carrier less than −70 dBc, which is good enough for our application.

### B. Noise in RF loopback

The primary performance characteristic of our system is the phase and amplitude noise power spectral density relative to the carrier for each individual readout tone. We measure this by connecting the output of the RF upconverter to the input of the RF downconverter via a 10 dB attenuator that mimics the transmission of the cryostat in a real experiment. For the first measurement we set the local oscillator to 5 GHz to access the 4–6 GHz RF band. We define 1000 equally spaced carrier frequencies within this band, whilst keeping at least 10 MHz away from the band edges and from the LO frequency to avoid filter roll-off effects in the IF chain. Carrier amplitudes are scaled such that the composite waveform fits in the DAC range. Attenuators in the RF electronics are set such that the downconverted waveform fits in the ADC range with sufficient overhead to avoid clipping. The demodulated complex responses of all carriers are then recorded for 400 seconds. This measurement produces 1000 separate complex waveforms, representing the amplitude and phase stability of the carriers. The amplitude and phase PSD of each waveform is obtained by calculating the FFT over a 100-second block of amplitude or phase response respectively, and averaging over 7 blocks with 50% overlap. We subsequently average the 1000 individual spectra to produce the uncorrected phase and amplitude power spectral density shown in Fig. 6a.

While the flat noise floor is due to the limited dynamic range of the ADC, the instability below 1 Hz is caused by overall drift of gain and phase in the electronics, which is to a large extent correlated. When operating the readout system on a detector array it is possible to remove correlated noise by inserting blind tones between the detector readout tones. The complex waveform of each detector tone is then divided, in the time domain, by the mean of the waveforms of two nearby blind tones. To test this method, two blind tones are assigned to each carrier, at a distance of 5 tone intervals on either side of the carrier. This corresponds to a MKID readout scheme where 10% of the carriers would be used as blind tones. The result is shown as the corrected plot in Fig. 6a. Blind-tone correction provides a strong reduction in noise below 1 Hz as it effectively removes correlated noise. The slightly increased

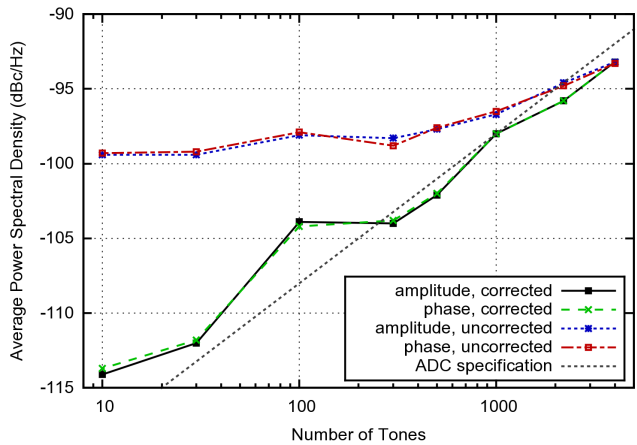

Fig. 7. Noise level in RF loopback (average over 0.1 - 80 Hz band, average over all carriers) as function of the number of multiplexed carriers. Also indicates the expected noise level based on ADC specifications, calculated as $-142$ dBc/Hz + $10 \times \log_{10}(n_{tones})$ + 14 dB (see also Table I).

noise above 10 Hz is due to the noise in the blind tones themselves.

We expect that the PSD relative to the carrier scales linearly with the number of tones: the noise level remains constant while the power per carrier decreases with increasing number of tones. The actual relation between number of tones and noise level is displayed in Fig. 7.

### C. Performance verification with an MKID array

The final performance evaluation consists of a read-out of a MKID detector array, where we use the full system as shown in Fig. 2. Fig. 8a shows the measured phase and amplitude noise of a lens-antenna coupled hybrid aluminium-NbTiN MKID, which is similar to the one described in [5]. The MKID is part of a 4x4 array, optimized for radiation detection at 850 GHz. The MKID has Q factor 125 000 and is read out with a $-91$ dBm carrier at 3.8 GHz, In the experiment we illuminate the MKID with a thermal radiation source at 6 K that couples $\sim$ 1 fW of radiation power to the device. More details can be found in [5] and [17].

We observe that the MKID amplitude noise is limited by the readout system noise if the number of readout tones > 100. Alternatively, the MKID phase noise is limited by the MKID intrinsic noise up to 4000 tones. Since both phase and amplitude noise result in the same detector sensitivity [17], the presented readout supports up to 4000 pixels without any degradation of device performance. Note that the intrinsic system noise, limited by the LNA, is lower as shown in Fig. 8b: The blue dotted line represents the measured noise PSD of to the LNA, the thin dotted line represents the digital readout performance (from Fig. 7) and the solid red line represent the expected noise PSD. The dots indicate the measured values based upon the blind tones not affected by MKIDs themselves.

### VII. CONCLUSION

We have developed a readout system for large arrays of microwave Kinetic Inductance Detectors, operating within a 2 GHz band in the 4–8 GHz range. Our system achieves a readout noise level of $-98$ dBc/Hz while reading 1000 carriers simultaneously at a rate of 159 frame/s. Blind tones are used to compensate for correlated drift. The noise of the readout system is far lower than the photon noise limit of NbTiN-Al MKIDs when using phase readout [5]. The system is scalable up to 4000 carriers, with a clear tradeoff between number of carriers and noise level. Similarly, the readout rate can be increased at the cost of decreasing the carrier frequency resolution.

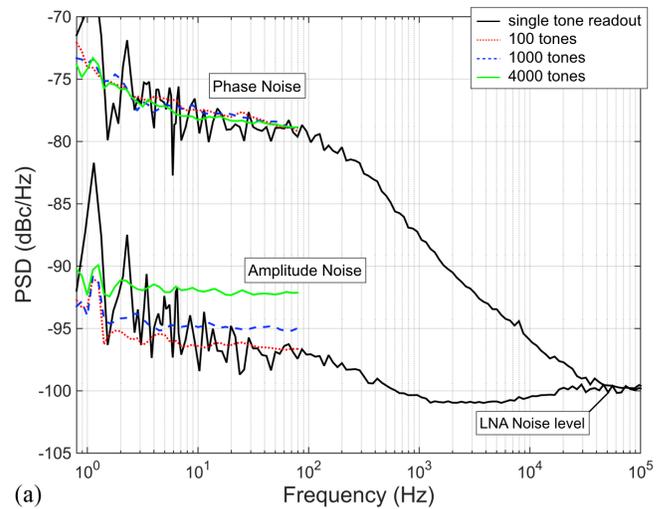

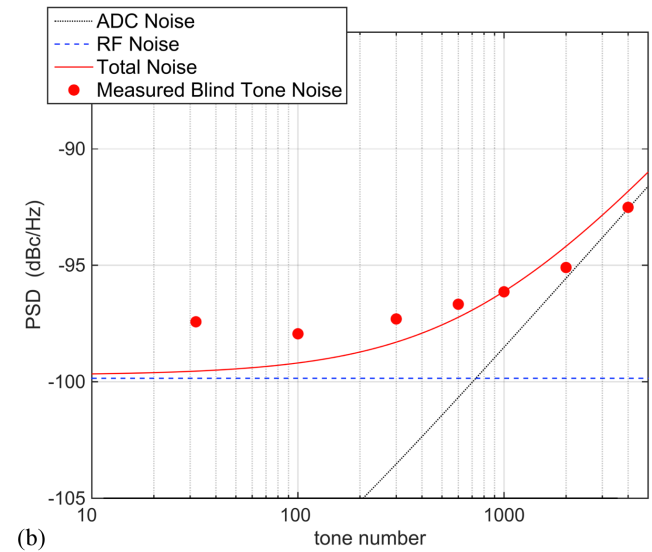

Fig. 8. (a) Measured system performance, including RF up- and downconverter on a MKID detecting $\sim$ 1 fW of radiation at 850 GHz, readout at 3.8 GHz. The solid lines represent the phase noise (upper line) and amplitude noise (lower line) of the MKID relative to the complex plane measured with a standard single tone homodyne readout. The shorter, dashed lines represent the performance of the readout presented here. (b) Total system noise performance as function of number of multiplexed tones, with contributions from ADC and RF electronics.